\documentclass[review,3p,times]{elsarticle}

\usepackage{ecrc}

\volume{00}

\firstpage{1}

\journalname{}

\runauth{J.Cho}

\jid{}

\jnltitlelogo{}

\usepackage{mathtools}
\usepackage{amsmath}

\usepackage{amssymb}
\usepackage{bm}

\usepackage[figuresright]{rotating}
\usepackage{caption}
\usepackage{subcaption}

\usepackage{multirow}
\usepackage{newunicodechar}

\usepackage[dvipsnames]{xcolor}
\newcommand{\red}[1]{ \textcolor{red}{#1}  }

\usepackage[utf8]{inputenc}
\usepackage{float}

\begin{document}

\begin{frontmatter}



\title{A Novel Mechanism for the Formation of Dislocation Cell Patterns in BCC Metal}

\author[a,b]{Jaehyun Cho}
\author[a]{Luke L. Hsiung} 
\author[a]{Robert E. Rudd}
\author[a]{Sylvie Aubry\corref{col1}}

\ead{sylvie.aubry@llnl.gov}
\cortext[cor1]{Corresponding author}

\address[a]{Materials Science Division, Physical and Life Sciences Directorate, Lawrence Livermore National Laboratory, Livermore, CA 94551, USA}
\address[b]{Analytical Mechanics Associates Inc., Thermal Protection Materials and Systems Branch, NASA Ames Research Center, Moffett Field, CA 94035, USA}

\dochead{}


\author{}

\address{}

\begin{abstract}
  In this study, we present the first simulation results of the formation of dislocation cell wall microstructures in tantalum subjected to shock loading.
  Dislocation patterns and cell wall formation are important to
  understanding the mechanical properties of the materials in which
  they spontaneously arise, and yet the processing and self-assembly
  mechanisms leading to their formation are poorly understood. 
  By employing transmission electron microscopy and discrete dislocation dynamics, 
  we propose a new mechanism involving coplanar dislocations and pseudo-dipole mixed dislocation arrays that is essential to the pattern formation process. 
  Our large-scale 3D DDD simulations demonstrate the self-organization of
  dislocation networks into cell walls in deformed BCC metal
  (tantalum) persisting at strain $\varepsilon=20\%$. 
    The simulation analysis captures several crucial aspects of how the dislocation cell pattern affects metal plasticity, as observed in experiments. Although experimental evidence is inconclusive regarding whether cell wall formation takes place at the shock front, after the shock, during release, or when the sample has had enough time to relax post-recovery, our simulations indicate cell wall formation occurs after the shock and before release. The extended Taylor hardening composite model effectively considers the non-uniform dislocation density when cell walls form and accurately describes the corresponding flow stress.
\end{abstract}



\end{frontmatter}

\section{Introduction}
Advanced metallic materials with superior mechanical properties enable the
ever greater structural performance needed across the spectrum
of modern technologies. Development of a material with increased
fracture toughness provides reliability and design flexibility of
devices in many fields such as transportation, advanced manufacturing,
and national defense. Mechanical properties of metallic materials are
strongly influenced by their crystallographic defect patterns called
microstructure. Among various microstructure types, formation and
evolution of dislocation structures have a pronounced impact on
mechanical properties of ductile metals such as flow stress. A
classical model describing the relationship between the flow stress
and dislocation structures is the Taylor hardening equation~\cite{widerisch_jm}:
\begin{equation}
  \tau = \alpha b \sqrt{\rho}
\label{eq:taylor}
\end{equation}
with $\tau$ the shear strength, $\alpha$ a dimensionless
material constant, $\mu$ the shear modulus, b the magnitude
of the Burgers vector, and $\rho$ the average dislocation density in the
metal. The basis for the power-law relationship in the Taylor
hardening law is the principle that the flow stress is set by the
spacing between obstacles to dislocation flow, which scales like
$1/\sqrt{\rho}$ assuming the dislocation density is uniform. However, the simple power-law dependence on $\rho$ may not be enough to describe how the strength of a metal depends on its dislocation-based microstructure. Many experiments~\cite{holt_jap, mughrabi_acta, kassner_acta} have endeavored to refine or redefine Taylor's law. One of the main drawbacks of Taylor's law is its loose dependence, solely contained in the parameter $\alpha$ and indirectly included in the dislocation density, on dislocation arrangements.



Dislocation cell structures are observed in plastic deformation experiments with wide ranges of materials and loading conditions~\cite{Keh_1963,RAJ1989233, mughrabi_mmta, devincre_science, guoetal_acta, haehner_acta, Gray_ARMR_2012}. Cell patterns are characterized by three-dimensional regions with dense dislocation entanglements (cell walls) along with less dense internal regions (cell interiors). The presence of walls of tangled dislocations reduces the
total elastic energy of the dislocation network by screening
long-range elastic fields from dislocations~\cite{holt_jap, moore_kwilsdorf_ss, kratochvil_rpa, mughrabi_jpa}. Important aspects of cell structure development in BCC metals found in experiments can be briefly
summarized as follows: (1) dislocation cell formation is frequently
observed with [001] crystal orientations~\cite{huang_hoansen_scripta,sauzay_kubin_pms}, (2) dislocation cell wall structures are typically observed in stage III and IV of work hardening~\cite{essmann_mughrabi_pm, rollet_kocks_msf}, (3) cell walls are composed of geometrically necessary dislocations (GNDs) which increases an average density of dislocations but does not alter a level of flow stress of the system~\cite{mughrabi_mmta, devincre_science, guoetal_acta, haehner_acta}, and (4) continuous development of dislocation cell structures eventually transforms into sub-grain structures bounded by sharp dislocation cell walls during dynamic recovery~\cite{hsiung_jpcm_j, bassiam_liu_fdi, hansen_barlow_book, Mustafa_mathon_baudin_msf}.

Many analytical models have been proposed to explain formation process
of dislocation cell structures~\cite{walgraef_jap, schiller_walgraef_acta, aoyagi_et_al_ijp, wu_prb, wu_zaiser_mt}.
These models concur that
uniform dislocation distributions become unstable for sufficiently
high dislocation density and form modulated dislocation structures
to reduce the elastic energy of the dislocations. By introducing
point-like dislocations with simplified slip systems and dynamics in 2D, cell pattern formation processes were investigated~\cite{Xia_msmse2015, madec_devincre_kubin_scripta, Kubin_book2013}. However, the 2D approach is limited in its ability to describe 3D dislocation
motion and interactions in complex network structures. Using 3D dislocation simulations, 
heterogeneous dislocation patterns in FCC metals in early-stage deformation (few percent of strain) have been studied~\cite{Xia_msmse2015, madec_devincre_kubin_scripta, GomezGarcia_PhysRevLett.96.125503}, and it was found that short-range interaction, e.g., cross-slip, plays a key role. Although the cross-slip mechanism may be
responsible for the early stage of dislocation patterning, it is questionable that dislocation core processes are still important for the formation of cell microstructures at larger strain (stage III and/or IV of work hardening) reported in experiments.

In this paper, we explain how dislocation cell walls are formed as
shown by Transmission Electron Microscopy (TEM) using 
discrete dislocation dynamics (DDD) simulation. First, we review TEM
analysis of dislocation cellular structures observed in deformed
tantalum, and then elucidate a new mechanism of cell wall formation
in BCC metals using DDD. The new mechanism is demonstrated in two steps. The
first crucial step is to obtain an initial dislocation network that is fully
relaxed using the theory of dislocations in an infinite
medium. The second step consists of loading the dislocation network
consisting of a fully relaxed configuration based on what we learn from step one but now using periodic boundary conditions (PBCs).
This procedure leads naturally to cell wall formation. Lastly, we validate our new mechanism by comparing it to
the features of cell walls exhibited in experiments.

\section*{High-explosive-shock-recovery experiment and TEM examination of dislocation structures}
\begin{figure}[htb]
  \centering \includegraphics[width=0.9\linewidth]{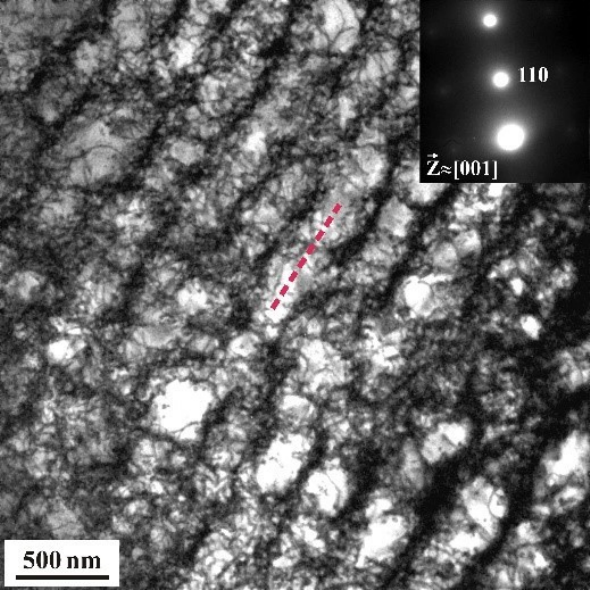}
  \caption{A well-defined dislocation substructure is observed in a shock-recovered tantalum specimen: a dislocation cell wall structure aligned with the $\langle$101$\rangle$ direction (red dotted line) in a [001]-oriented grain. The shock direction [001] is into the plane of the page. This TEM micrograph was acquired from a $\sim$200nm thick sample in a Transmission Electron  Microscope operating at 200 keV.} 
   \label{fig:tem}
\end{figure}

Ingot metallurgy (IM) tantalum samples (commercially pure tantalum) in
the form of plate stock produced using a standard electron-beam
melting process were obtained from Cabot Corporation, Boyertown,
PA. Details of high-explosive-driven (HE-driven) hubcap
shock-recovery experiments employed for the current investigation can
be found in~\cite{campbell_et_al_jap}. Briefly, one single explosively driven shock-recovery experiment was conducted by detonating explosive on
hubcap alloy plates (3mm thick), which were shocked into
polyurethane foams immersed in a water tank. The shock experiments
were carried out under a peak pressure of $\sim$30GPa, as simulated
using a CALE continuum hydrodynamic code~\cite{barton_na}. As shown
in Figure~\ref{fig:tem}, a dislocation cell wall structure is
observed in a grain oriented with the loading axis parallel to the
[001] orientation with a mean dislocation density estimated to be around
$\bar{\rho}\sim{10}^{14}/{\mathrm{m}}^2$. The cell-wall structure tends
to align parallel to the projected $\langle$101$\rangle$ directions (shown as
red dotted line) that have an angle of 35.16$^{o}$ with respect to the
adjacent $\langle$111$\rangle$ Burgers vectors.
 
\section{Pseudo-dipole dislocation cell structures~\label{pseudo}}
A careful dislocation analysis of Figure~\ref{fig:tem} done along the $[001]$ view direction, $\langle101\rangle$ direction reveals that each pair of Burgers vectors belonging to the $\{111\}$ family shares the same plane. More precisely, ${\bf b}_1=\pm[111]$ and ${\bf b}_2=\pm[\bar{1}1\bar{1}]$ share the slip plane ${\bf n}_1=(\bar{1}01)$ plane, and ${\bf b}_3=\pm[\bar{1}\bar{1}1]$ and ${\bf b}_4=\pm[1\bar{1}\bar{1}]$ lie together on the same ${\bf n}_2=(101)$ slip plane~\cite{hsiung_llnl_internal}. We define this pair of Burgers vectors, sharing one slip plane as {\it{coplanar slip systems}}.

\begin{figure}[htb]
  \centering \includegraphics[width=\linewidth]{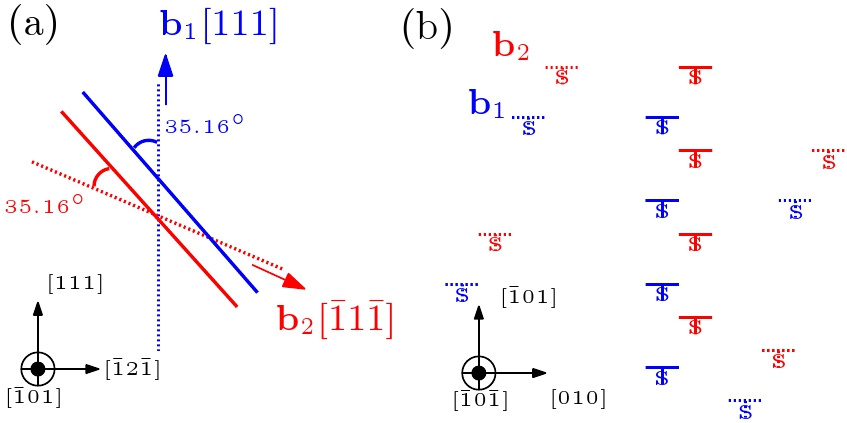}
  \caption{Schematic example of the two key mechanisms of cellular structure formation: (a) rotation of co-planar screw dislocations (${\bf b}_1$ and ${\bf b}_2$ -- dotted lines) to form a pseudo-dipole mixed dislocation pair (solid lines), and (b) clustering of pseudo-dipole mixed dislocations pairs via elastic interactions from random (dotted) to wall (solid) configurations.}
   \label{fig:coplanar}
\end{figure}

A novel mechanism stemming from the coupling reactions of dislocations lying on coplanar slip systems was found to explain the formation of low-energy-type dislocation substructure as shown in Figure~\ref{fig:tem}, as described in an unpublished report~\cite{hsiung_llnl_internal}. 
The mechanism involves the following hypotheses: 
(I) In the early deformation stage, the dislocation network is mainly composed of screw dislocations.
(II) Each pair of screw dislocations lying on coplanar slip systems - ${\bf b}_1$ and ${\bf b}_2$ / ${\bf b}_3$ and ${\bf b}_4$ - alters their line directions to be parallel to the [101] direction on their slip plane - ${\bf n}_1$ / ${\bf n}_2$, respectively - through elastic interactions, forming mixed dislocation pairs.  Figure~\ref{fig:coplanar} (a) shows an example of the coupling reaction between two screw dislocations on coplanar slip systems (dashed lines) resulting in their re-alignment from screw to mixed dislocations.
More precisely, one dislocation of the pair turns counterclockwise and the other turns clockwise, both by an angle of 35.16$^{\circ}$. Once aligned, each pair includes two different Burgers vectors that attract each other forming a pair. The formed configuration is referred as a {\it pseudo-dipole}. Our definition of a pseudo-dipole is different than a dipole defined as a pair of aligned dislocations with equal and opposite signs of Burgers vectors. (III). Pseudo-dipole arrays of dislocations, composed of many pseudo-dipoles, form a relaxed microstructure - resulting in stress-screening dislocation arrays. 
Fig~\ref{fig:tem} (b) shows an example of the relaxed pseudo-dipole array structure composed of ${\bf b}_1$ and ${\bf b}_2$. The pseudo-dipole array is a different low energy dislocation structure than Taylor lattice~\cite{NEUMANN1986465} where pairs of aligned dipoles with a Burgers vector. (IV) Under external loading, these pseudo-dipoles travel and form locked configurations - stable cell walls - determined by the long-range stress fields and short-range dislocation core interactions e.g., junction formation. 

The hypotheses describing the newly introduced mechanism are investigated and confirmed in the next two sections using DDD.

%



\section{Formation of mixed dislocation pseudo-dipoles}\label{relaxation}

The formation of pseudo-dipole arrays as described in the previous section is investigated in this section using the DDD computer code ParaDiS~\cite{arsenlis_msmse}.
To allow for a relaxation of dislocation networks composed of an initial array of screw dislocations followed by rotations rotations on coplanar slip systems without changing their lengths, a cylindrical-shaped simulation domain is used.
\begin{figure}[htb]
  \centering \includegraphics[width=\linewidth]{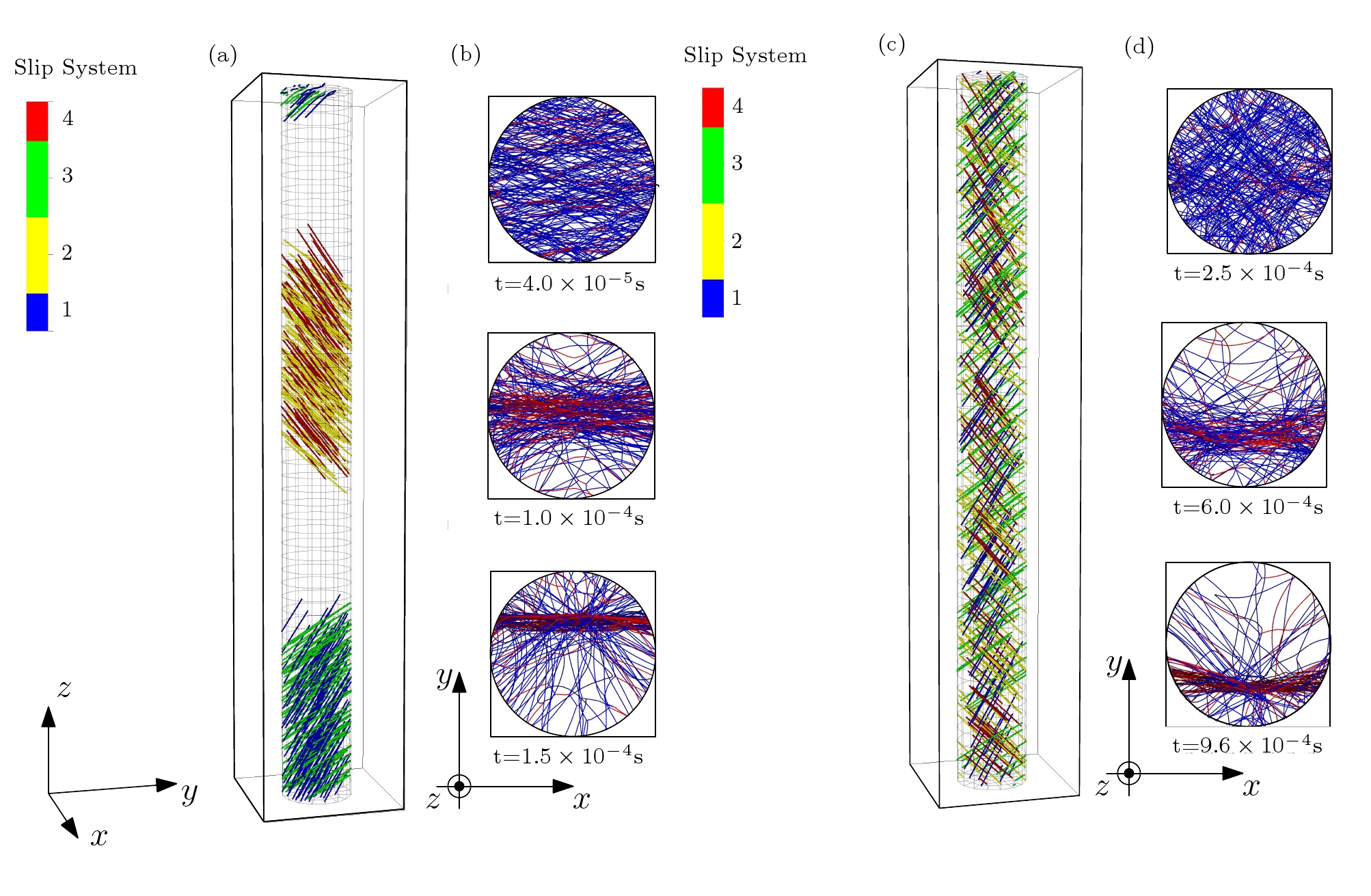}
  \caption{(a) and (c): Two different initial distributions of screw dislocation pairs: randomly spread over two local regions and over an entire domain, i.e., the "local-random system" and the "full-random system", respectively; (b) and (d): Snapshots of relaxation of the two systems (a) and (c), respectively. The blue and red lines represent glissile and junction dislocations, respectively.}
   \label{fig:mixed}
\end{figure}
Figure~\ref{fig:mixed} (a) and (c) show the set-up of two simulations comprising a cylindrical system with radius 2,000${\bf b}$ and height 30,000${\bf b}$. In the first system (a), 400 screw dislocations are positioned in the
domain with gaps as follows: 
first, 200 screw dislocations with Burgers
vectors (${\bf b}_1$ and ${\bf b}_2$) are randomly populated
in -15,000$|{\bf b}|\leq z <-7,500|{\bf b}|$, where $z=[001]$ is the longitudinal coordinate.
The remaining 200 screw dislocations with the other slip systems (${\bf b}_3$ and ${\bf b}_4$) are
arbitrary inserted in the range $0 \leq z < 7,500{\bf b}$.
The constructed sets of co-planar slip systems
are separated by a distance of 7,500${\bf b}$ which is
sufficiently large for a set of dislocations in each coplanar slip system to
closely interact with each other at the beginning of the simulation
without interference from dislocations of non-coplanar dislocations. In the second system (c), 400 screw dislocations are randomly inserted over the whole cylinder domain.
In the following, we refer to these two systems (a) and (c) as
“local-random system” and “full-random system”, respectively.  
The local- and full-random systems are relaxed assuming they are composed of  infinite dislocations located in an infinite medium and under zero external loading to find their low energy dislocation structures.

In order to allow dislocations to move freely in the simulation domains,
dislocations are modeled as semi-infinite dislocations: radial
surfaces of the cylinders are treated as infinite boundaries, and the
vertical plane ($z$) is handled using PBC. The
contributions of quantities such as force, stress and velocity field, calculated in dislocation
dynamics, are modified by implicitly extending dislocations to
infinity using virtual straight segments~\cite{weinberger_msmse}.
With this mixed boundary condition, we can expect each dislocation to travel as one infinitely across the $x$-$y$ plane and periodically in the $z$ planes of the end caps.


Figure~\ref{fig:mixed} (b) and (d) show snapshots of the relaxed configurations for
the two systems in top view along the $[001]$ axis at a series of times. The blue and
red lines represent glissile and junction dislocations
respectively. Both systems exhibit similar relaxation progress: the
initial screw dislocations rotate toward their  mixed character
and naturally align along $\langle101\rangle$ direction forming pseudo-dipole arrays.
The rotated dislocations entangle and form compact structures of width $\sim 600{\bf b}$
($\sim$180~nm) in the $y$ direction. During relaxation, the
total density of dislocations decays by about 50$\%$ to release the
elastic energy stored in the dislocation network and saturates at a
constant dislocation density level. The local-random system undergoes a faster relaxation toward equilibrium compared to
the full-random system. This rapid relaxation to a low
energy state is due to the rotation of the screws to mixed character occurring only between the sets of coplanar screw dislocations screw dislocations lying in coplanar slip systems. More precisely, in the partial-random system, each
of the two groups of screw dislocations,
(${\bf b}_1$ and ${\bf b}_2$) and (${\bf b}_3$ and ${\bf b}_4$),
located at the top and bottom part of the cylinder, aligns and forms
a compact structure. In the case of the full-random system, initial
screw dislocations rotate at a lower rate due to the adjacent interactions with non-coplanar slip systems, but eventually this system also
ends up relaxing to a similar microstructure as the local-random
system. These simulations prove that independently of the initial configuration, the relaxation of screw dislocations lying in coplanar slip system give rise to bundle of dislocations which form an inhomogeneous dislocation microstructure and confirm the hypotheses I, II, and III described in Section~\ref{pseudo}.

\section*{Formation of dislocation cell walls during plastic deformation of Ta under shock loading}
In Section~\ref{relaxation}, we validated our hypothesis that an initial screw-dislocation-dominated network relaxes to form pseudo dipole arrays, where dislocations rearrange along the $\langle101\rangle$ direction and group during relaxation as suggested by experimental data~\cite{hsiung_jpcm_j,hsiung_llnl_internal}. Now, we investigate whether an initial $\langle101\rangle$ mixed dislocation array can yield a dislocation cell structure during loading as observed in the TEM for tantalum shown in Figure~\ref{fig:tem}. 

A simulation box of dimension 300nm$\times$300nm$\times$300nm (300nm$\approx$1000$\bf{b}$) containing an initial array
of mixed dislocations with arbitrary Burgers vectors aligned in the $\langle101\rangle$ direction, or pseudo-dipole array, is randomly populated up to an initial dislocation density of 0.75$\times{10}^{14}/{\mathrm{m}}^2$. In the following, we refer to this simulation set up as "pseudo dipole-array". A loading stress is
applied to the system at a strain rate of $\dot{\varepsilon}={10}^5/{\mathrm{s}}$
along the $[001]$ axis. A counterpart simulation we will refer to as a "screw-array" is initiated with a screw dislocation array randomly populated at the density similar to the initial pseudo dipole array. In contrast to the relaxation simulations executed in the previous section, the strain-rate controlled simulations are carried out using PBCs in all directions.

\begin{figure*}
  \centering \includegraphics[width=\linewidth]{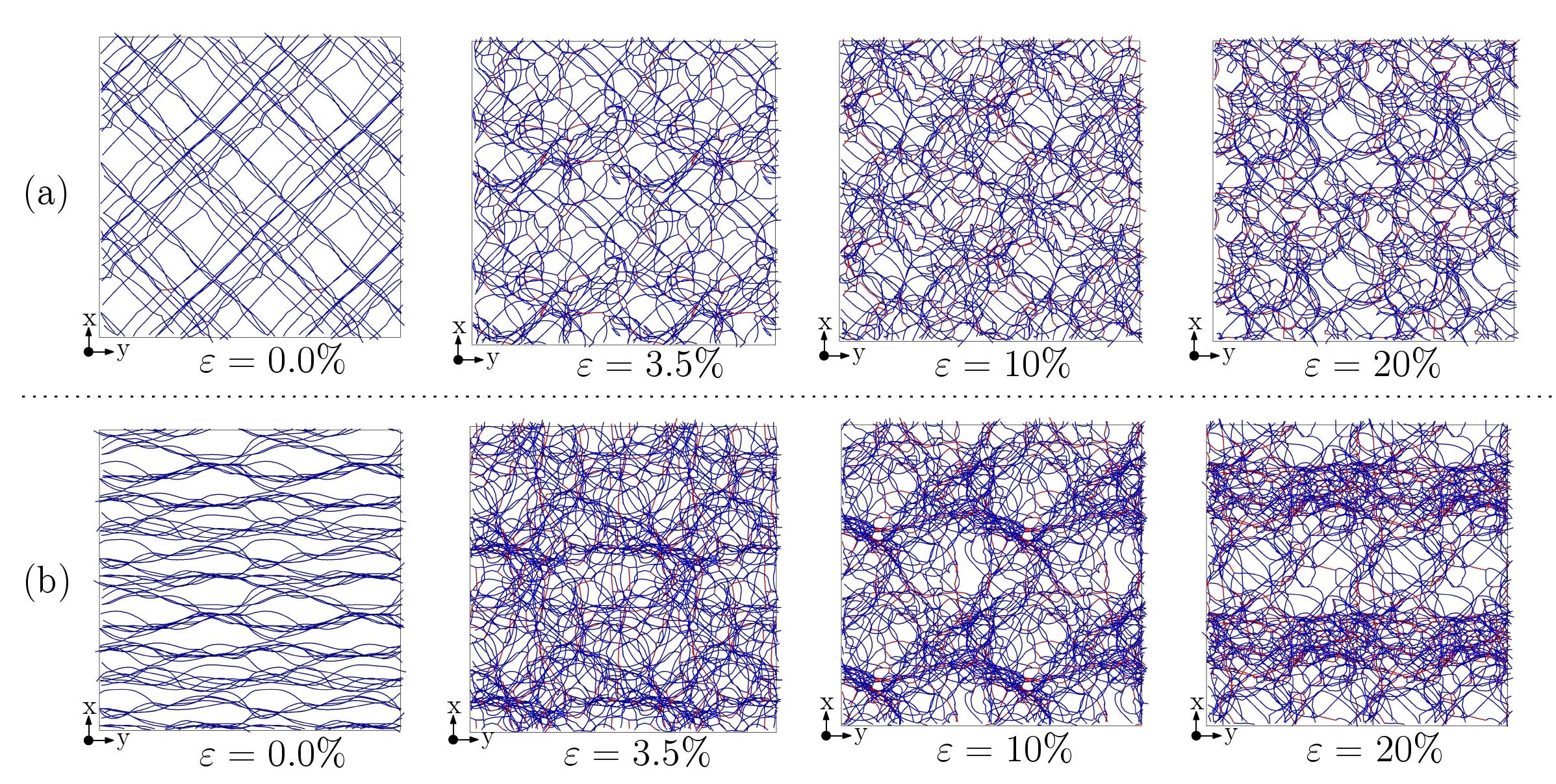}
  \caption{Snapshots of dislocation networks shown viewed along [001] at 0$\%$, 3.5$\%$, 10$\%$ and 20$\%$ of total strain under [001] uniaxial deformation at a strain rate of $\dot{\varepsilon}={10}^5/{\mathrm{s}}$: (a) The domain is initially populated with screw dislocations; (b) The domain is initially populated with mixed dislocations or pseudo dipole array. For better visual understanding of cell wall structure, the domain box (L=300nm$\approx1000\bf{b}$) is periodically replicated along the $x$ and $y$ directions (2$\times$2). Glissile and junction dislocations are shown in blue and red segments, respectively. The simulation video of cell formation corresponding to (b) can be found in the supplemental material.}
   \label{fig:simulation}
\end{figure*}

Figure~\ref{fig:simulation} shows snapshots of the two DDD simulations in viewed along [001] at strains $\varepsilon$=0$\%$, 3.5$\%$, 10$\%$, and
20$\%$ respectively. Figure~\ref{fig:simulation} (a) shows the simulations results starting from the screw array and (b) starting from the pseudo dipole array. 
Dislocation lines are colored by the Burgers vector types: blue and red represent glissile and junction Burgers vectors, respectively. For a better visual understanding of the dislocation microstructures, the periodic unit domain is replicated one time in each direction $x$ and $y$. 
The two simulations exhibit different dislocation structures after a few percent strain is reached. 
As the strain increases in each simulation, the screw array evolves to a network with a homogeneous distribution of dislocations~\footnote{\label{fnote1}Evolution of pair correlation analysis of dislocation microstructures of the two systems with the screw array and the pseudo array can be found in the supplemental material.} composed of glissile and junction dislocations at $87\%$ and $13\%$ percentages, respectively. However, the pseudo dipole array exhibits a microstructure transitioning from a homogeneous dislocation distributions observed between 0$\%$ and 3.5$\%$ strain to cell walls at
strain levels between 10$\%$ and 20$\%$. 
More precisely, at 10$\%$ of strain, the cell walls are loosely equiaxed in $x$ and $y$ axes loosely forming two-dimensional cellular structures. As strain continues further, such two-dimensional cell structures evolve and change their shapes and positions. As the strain approaches to $20\%$, the cell walls become one-dimensional structures distinctly extended in $y$ axis and non-evolving. In the cell interiors, dislocation networks are composed of $\sim13\%$ junctions as the simulation of the screw array, while the cell walls contain $\sim27\%$ junction dislocations. Such a high density of junctions in the clusters helps to keep the quasi-static cell wall structures without changes of shapes and positions. This observation confirms hypotheses IV described in Section~\ref{pseudo}. In the supplemental material, we discuss in more detail about dislocation structures of the simulations. In the following discussion, we term these two systems where the cell walls are formed and the homogeneous distribution of dislocations is observed as “cell-wall system” and “no-cell system”, respectively.

\begin{figure*}[htp]
  \centering \includegraphics[width=0.8\linewidth]{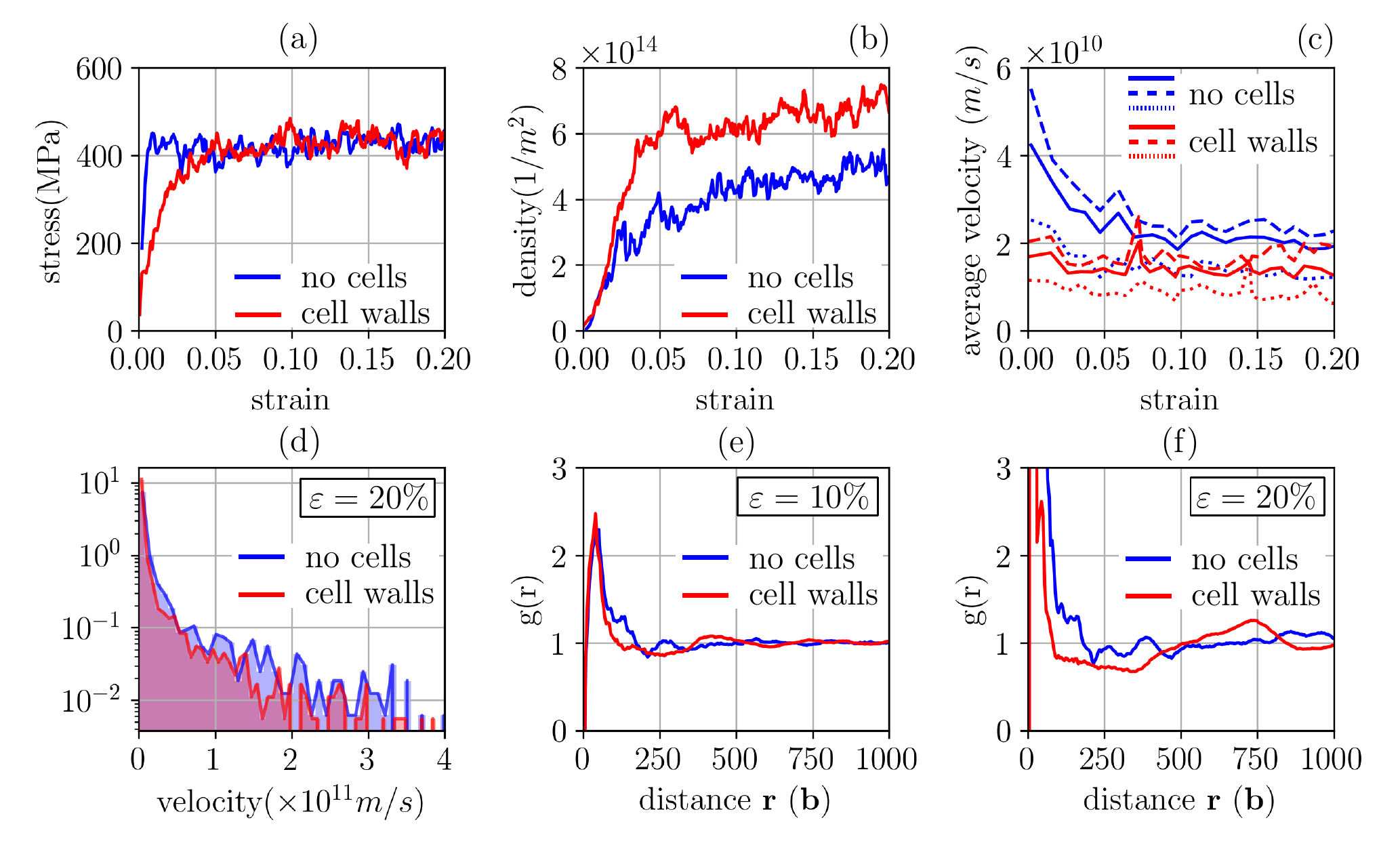}
  \caption{Analysis of two dislocation dynamics simulations which
    differ in their initial populations: screw dislocations (blue
    curves) and mixed dislocations (red curves) arrays, where
    homogeneous (no cells) and heterogeneous (cell walls) dislocation
    distributions are observed, respectively. (a) stress-strain
    relationships, (b) evolution of the dislocation densities, (c)
    evolution of average velocities (solid lines), in the tangled
    regions (dotted lines) and in the spatial areas (dashed lines),
    (d) probabilities of dislocation velocities at 20$\%$ strain and
    (e) and (f) dislocation pair correlation function of the
    dislocation distributions at 10$\%$ and 20$\%$ strains,
    respectively.}
   \label{fig:analysis}
\end{figure*}

Figure~\ref{fig:analysis} (a) shows the stress-strain response of the two systems. Both systems saturate at about the same flow stress level of 410~MPa. Note that the cell-wall system's stress saturates at $\sim$5\% strain, at which point the cell patterning has begun but well-defined cell walls form later at $\sim$10\% strain (cf.~Figure~\ref{fig:simulation} (b)). Evolution of the corresponding dislocation densities is shown
in Figure~\ref{fig:analysis} (b). The cell-wall system saturates at $\sim$1.5 times
higher dislocation density (0.7$\times{10}^{14}/{\mathrm{m}}^2$) than the
no-cell system (0.47$\times{10}^{14}/{\mathrm{m}}^2$). This observation reveals
that the presence of dislocation cell structure increases dislocation
density levels without increasing the corresponding flow stresses. 

To understand the density increase in the presence of the cell walls,
we analyze the obtained microstructures by distinguishing
between regions with high and low local density using the kernel density estimation (KDE) method. We will call high- and low-density regions as "knotted" and "unknotted" in the following. 
Figure~\ref{fig:analysis} (c) shows the time evolution of the dislocation velocities averaged over different sets of dislocations: all of them (solid lines), knotted (dotted lines), and unknotted (dashed lines) for each of the two systems. We find that the cell-wall system exhibits overall slower dislocations than the no-cell system by $\sim$67\% indicating that the cell walls reduce the average dislocation velocities by impeding dislocation motion, and the differences of average velocities and the dislocation densities yield the same constant strain rate according to Eshelby's equation, where the strain rate is given by
\begin{equation}
  \dot{\varepsilon}=\frac{1}{M}b\overline{v}\rho
\end{equation}
where $M$ the orientation factor, b the magnitude
of the Burgers vector, $\rho$ the dislocation density, and $\overline{v}$ the average dislocation velocity.  

Moreover, dislocations in knotted regions of the cell wall system (red dotted line) move much more slowly than the other kinds of dislocations, 
meaning that the obtained cell wall structures are more stable 
than any microstructures where cell walls are not present. 
Interestingly, as the the cell walls contain more and more dislocations starting at $\sim12.5\%$ strain, the velocity of dislocations in the unknotted (red dashed line) increases. Figure~\ref{fig:analysis} (d) shows the probability of dislocation
velocities of the two systems at 20$\%$ strain. These results confirm that
dislocations located in the cell walls move slower, and some not at all, in the cell-wall system than in the no-cell system.  

Figure~\ref{fig:analysis} (e) and (f) shows pair correlation analysis of dislocation networks of the two systems at 10$\%$ and 20$\%$ strain levels, respectively. Correlation analysis result shows a signal starting at around $500 \bf{b}$ with the distance 300$\bf{b}$ indicating the presence of the wall structure and how it becomes more pronounced after 10$\%$ of strain.


According to Mughrabi~\cite{mughrabi_mmta}, the $\alpha$-factor of the Taylor hardening equation (Equation 1) decreases when dislocation cells are present during
steady-state deformation, which is also observed in our
simulations: $\alpha$-factor is reduced by 8.5$\%$ when cells are present compared to when they are not. Since cells walls create more dislocation density which little affects of the stress response of the material when compared to dislocations homogeneously distributed over the domain~\cite{gao_huang_nix_hutchinson_jmps, gao_huang_scripta}, the Taylor hardening model needs to be refined 
to consider a different strengthening effect for a given dislocation density. 

Mughrabi~\cite{mughrabi_acta, mughrabi_mmta, mughrabi_jpa} observed dislocation cell formations under cyclic loading, and characterized the hardening effect using a composite model~\cite{mughrabi_jpa, servillano}:
\begin{eqnarray}
  \sigma&=&f_{\mbox{wall}}\sigma_{\mbox{wall}}\ +\ f_{\mbox{interior}}\sigma_{\mbox{interior}} \nonumber \\
  &=& \alpha\mu {\bf b}\left(f_{\mbox{wall}}\sqrt{\rho_{\mbox{wall}}}\ +\ f_{\mbox{interior}}\sqrt{\rho_{\mbox{interior}}}\right)
\label{eq:composite}
\end{eqnarray}
where $f_{\mbox{wall}}$ ($\rho_{\mbox{wall}}$) and $f_{\mbox{interior}}$ ($\rho_{\mbox{interior}}$) are the volume fractions (dislocation densities) of the high
and low density regions, respectively. 

\begin{figure*}[htb]
  \centering \includegraphics[width=\linewidth]{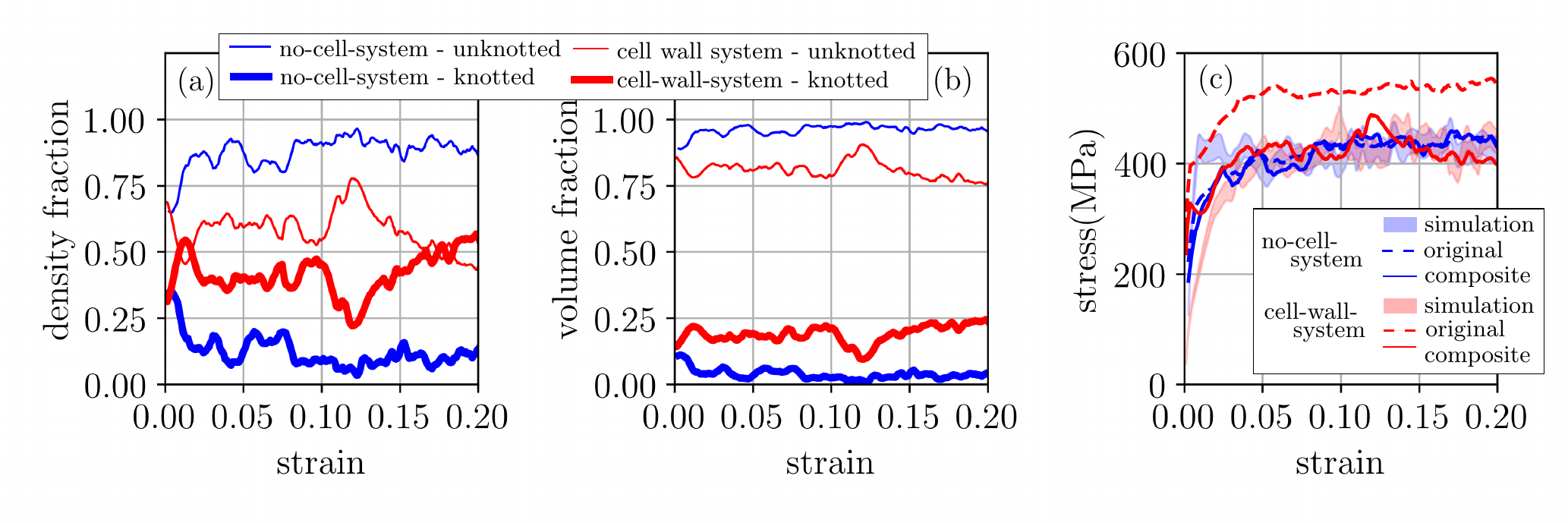}
  \caption{Predictions of flow stress using Taylor's law for the no-cell and cell-wall systems, indicated with blue and red colors, respectively: (a) of dislocation density fractions in high-density ($\rho_{\rm{wall}}/\rho$)
regions (knotted), and low-density ($\rho_{\rm{interior}}/\rho$) regions (unknotted) with thick and thin solid lines, respectively; (b) Evolution of volume fractions of high ($\rm{Volume}_{\rm{wall}}/{\rm{Volume}}$) and low ($\rm{Volume}_{\rm{interior}}/{\rm{Volume}}$) regions (knotted and unknotted) are shown with thick and thin solid lines, respectively; (c) Predictions of the original Taylor and composite models using DDD simulations, are shown dashed and solid lines, respectively. For no-cell-wall system, both original and composite models successfully predict as shown in blue dashed and blue solid lines, respectively. For  cell-wall system, the original model is in poor agreement with Taylor's law while the composite successfully predicts the numerical data as shown with the red dashed and red solid lines, respectively.}
   \label{fig:composite}
\end{figure*}

Similarly to the previous analysis, the two high- and low-density regions can be defined as knotted and unknotted.
Evolution of the knotted and unknotted populations of dislocations are shown in Figure~\ref{fig:composite} (a) and (b) as characterized by density and volume fractions, respectively.\footnote{The threshold we employ to separate the high-density and low-density regions is $1.3\times10^{-9}/m^2$.} We find that the presence of a cell wall leads to a system in which $\sim$20\% of the volume with knotted dislocations contains almost the same amount of dislocations as the remaining $\sim$80\% of volume with unknotted dislocations. Without the presence of the cell wall (blue curves), dislocations are quasi-uniformly distributed over the domain, and the system is mostly  unknotted dislocations.  

In Figure~\ref{fig:composite} (c), simulation results of the cell-wall system and no-cell system are shown in blue and red areas, respectively, along with fits to the results using the Taylor model (Equation~\ref{eq:taylor}) and the composite model (Equation~\ref{eq:composite}) as shown by dashed and solid lines, respectively, in the corresponding color. The constant $\alpha$ is taken to be 1.3 for all predictions. We find the original Taylor model fits the quasi-uniformly distributed data correctly, but fails to describe the hardening when heterogeneous dislocation clusters are present (red dashed line). The composite model of flow stress is in good agreement with both systems and successfully predicts strain hardening effects.

\section*{Discussion}
In summary, we verified the key mechanism leading to dislocation cell formation, and successfully proved its validity by showing spontaneous formation of cell walls using the DDD approach and analysis. The mechanism is
based on the reaction between co-planar dislocations and results in alignment and clustering of $\langle101\rangle$ mixed dislocation
arrays: pseudo-dipole dislocation arrays. When the domain is seeded with  $\langle101\rangle$ mixed dislocations and is subjected to constant strain-rate loading, 
dislocation cell patterns spontaneously form and become more pronounced as plastic strain increases to $\sim20\%$ (stage III/IV of work hardening).
The presence of cell walls lead to an additional accumulation of dislocations while maintaining the same level of saturated flow stress as in a quasi-uniform dislocation distribution. 
We found the composite model is in good agreement with the flow stress for systems both with and without cell walls.

Several models in the literature~\cite{mughrabi_acta,wu_zaiser_mt,zhou_dynamic_2015} predicting dislocation cell patterns have focused on elastic interactions between dipoles of straight dislocations including Burgers vectors of equal and opposite signs.
In this study, we found a new assembled configuration of low energy dislocation structures composed of dislocations with Burgers vectors that are not strictly opposite, but have a significant anti-parallel component: pseudo-dipole mixed dislocations, which turned out to be the key ingredient of dislocation cell structures.



Determining the timing and process of cell wall formation in shock experiments is a challenging task. It is unclear whether the cell walls develop during the initial shock rise at extremely high strain rates ($\sim{10}^7$/s), during the post-shock plastic flow, or after the release. Our simulations indicate that cell walls are only visible at intermediate strain rates ranging from ${10}^4$/s to ${10}^5$/s, similar to the post-shock stage where pressure fields persist during the tens of microseconds pulse duration in explosively driven flyer-plate experiments~\cite{meyer_ds}. We did not observe any cellular dislocation structures forming at high strain rates between ${10}^6$-${10}^7$/s, which is comparable to the 30~GPa range of the initial shock rise.

Additionally, our simulation indicates that initial dislocations should relax in an energetically favorable pseudo-dipole configuration before reaching the post-shock stage. Although we did not examine the shock release in our DDD simulation, we anticipate that the resulting cell walls are stable enough to survive the shock release and be observed under ambient conditions, as demonstrated in TEM observations (Figure~\ref{fig:tem}).


The objective of this paper is to emphasize the novel process of cell creation through the utilization of dislocation dynamics simulations. There are various possibilities for expansion of this research. First, a bigger simulation with an increased amount of initial pseudo-dipoles may be explored to model more intricate cellular configurations. Second, material parameters such as Poisson ratio and shear modulus, which are dependent on pressure and temperature, can be enhanced~\cite{Moriarty,Orlikowski} to more accurately simulate high-explosive-shock-recovery experiments in a quantitative manner.

\section{Acknowledgements}
This work was performed under the auspices of the U.S. Department of
Energy by Lawrence Livermore National Laboratory (LLNL) under Contract
DE-AC52-07NA27344 and was supported by the LLNL Laboratory-Directed Research and Development Program under Project No. 20-LW-027. Release number: LLNL-JRNL-829087. 


\bibliographystyle{elsarticle-num}
\bibliography{mybib}

\newpage

\appendix

\renewcommand{\figurename}{A. Fig.}

\section*{Appendix 1: Coupling reaction of a pair of co-planar dislocations}
\begin{figure}[h!]
  \centering \includegraphics[width=\linewidth]{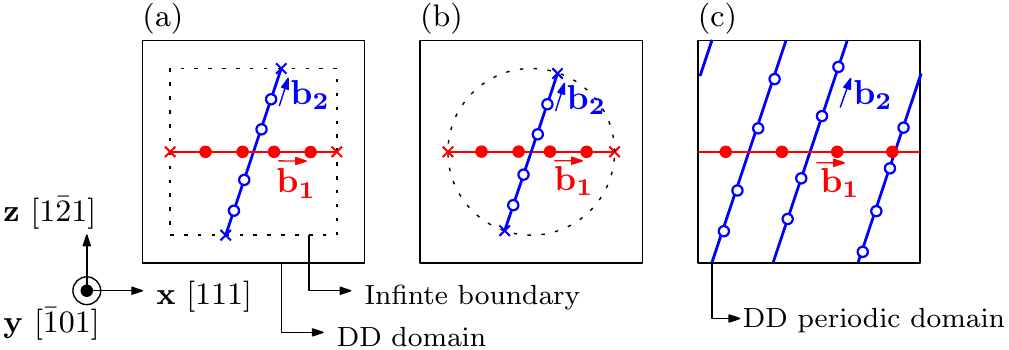}
  \caption{A pair of screw dislocations (${\bf b}_1$=[111] and
    ${\bf b}_2$=[$\bar{1}0\bar{1}$]) separated by
    200$|{\bf b}|$ in ${\bf y}$-axis subjected to three
    different boundary conditions: (a) cuboid with free surfaces, (b)
    cylindrical domain with free surfaces and (c) cuboid box with
    periodic boundary conditions.}
   \label{fig:pair}
\end{figure}

In A. Figure~\ref{fig:pair}, pairs of screw dislocations
${\bf b}_1$=[111] and ${\bf b}_2$=[$\bar{1}1\bar{1}$]
are inserted and subjected to the three different boundary conditions: (a) cuboid box
with free surfaces, (b) cylinderical domain free surface and (c)
cuboid box with periodic boundary conditions (PBCs) as seen in A. Figure~\ref{fig:pair}.
Two screw dislocations are separated by 200$|{\bf b}|$ in the
y=$[1\bar{10}]$ axis. In the case of free boundary conditions, the
straight dislocations are terminated at the domain boundaries, and the
two screw dislocations can cross each other once at the center of the
system. While in the case of the PBC, one end of the straight
dislocation is continuously extended through the opposite side of the
system leading to multiple periodic images in the simulation
domain. Next, the system is relaxed.

\begin{figure}[h!]
  \centering \includegraphics[width=\linewidth]{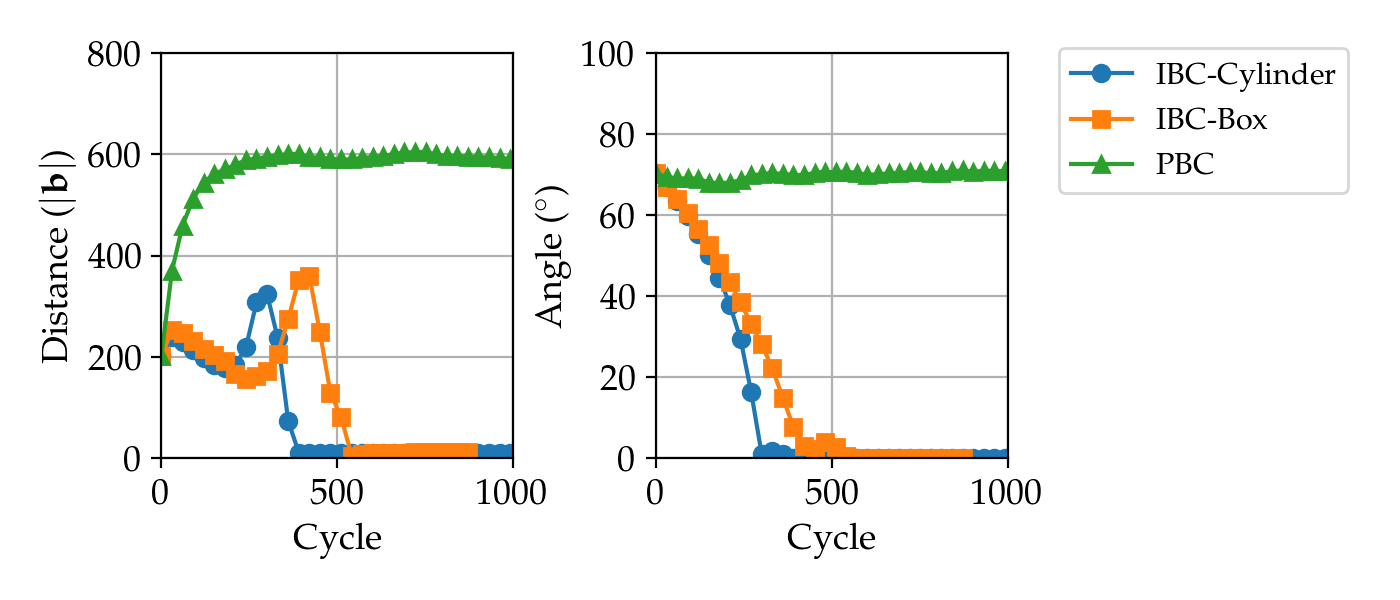}
  \caption{Relaxation results of an initial pair of screw dislocations
    (${\bf b}_1$=$[111]$ and
    ${\bf b}_2$=$[\bar{1}1\bar{1}]$) separated by
    200$|{\bf b}|$ in the ${\bf y}$-axis subjected to three
    different boundary conditions: (a) cuboid free surfaces, (b)
    cylindrical domain with free surfaces and (c) cuboid periodic
    box.}
   \label{fig:pair_result}
\end{figure}

A. Figure~\ref{fig:pair_result} shows relaxation results: (a) evolution of the shortest
distances between screw dislocation pairs and (b) evolution of angles
between screw dislocation pairs.  In the IBC case, the screw
dislocations approach each other and rotate to align with each
other. More precisely, each of two screw dislocations rotates in
opposite direction to reduce the intersection angle from 71.2$^{o}$ to zero
degree. As they align, the distances between the pairs slightly
increase and then decrease to zero. The two free surface boundary
cases lead to different dislocation behaviors, but eventually ends up
at the same result.  With PBC, the rotation and alignment are not
observed, the two dislocations move apart from each other in the y
direction to reduce the elastic energy between the screws.  The
elastic interaction between the screw dislocations yields different
rotating direction, and  the rotation and alignment observed in the
infinite boundary cases.

\begin{figure}[h!]
  \centering \includegraphics[width=0.9\linewidth]{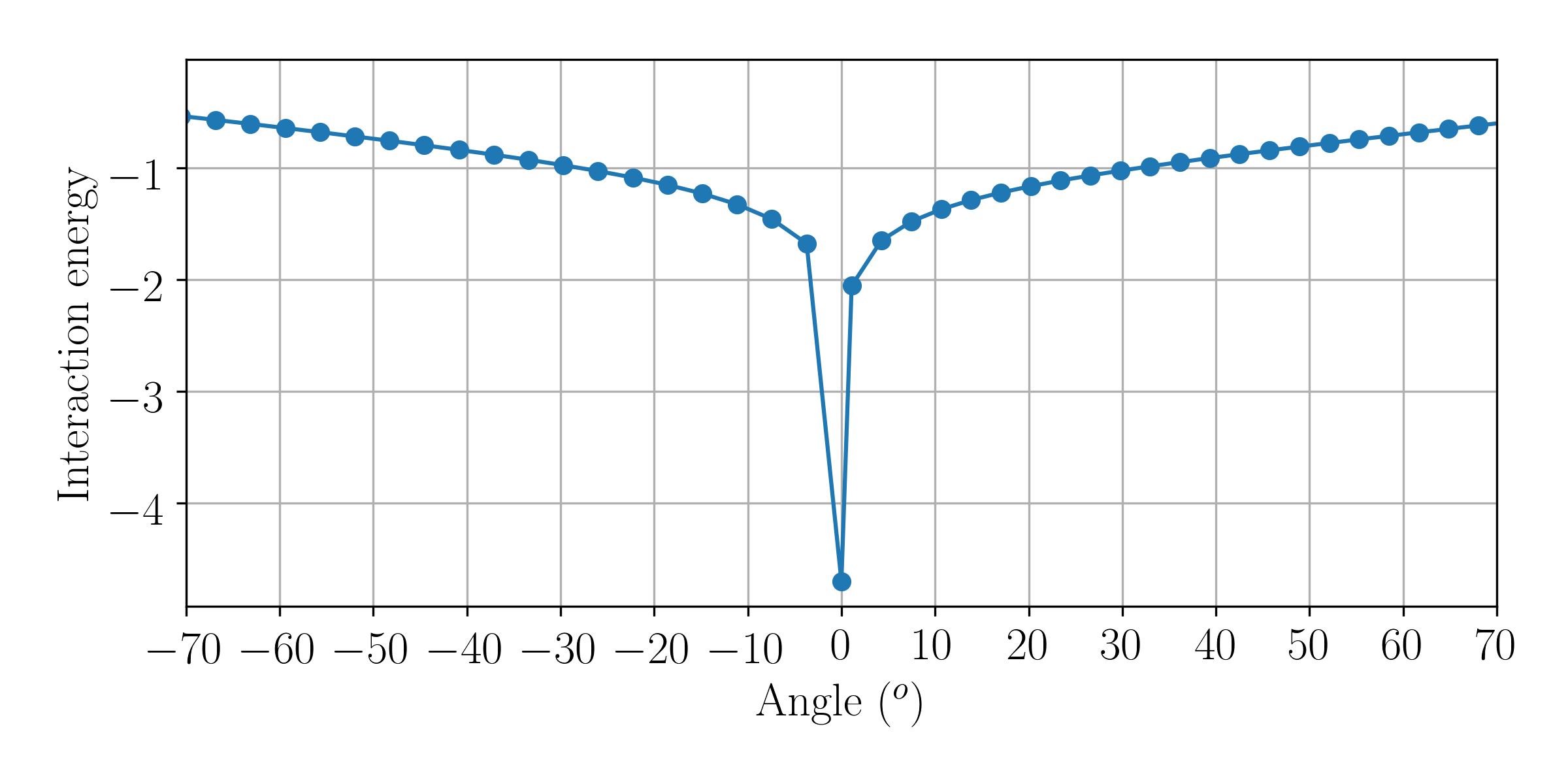}
    \caption{The variation of elastic interaction energy between two straight dislocations as a function of the angle between them.}
    \label{fig:rotE}
\end{figure}

A. Figure~\ref{fig:rotE} shows the variation of interaction energy~\cite{hirth1992theory} between two straight dislocations as function of intersection angle.
Character angles of the two straight dislocations are screws forming the intersection angle $70^{\circ}$ degree. 
As the dislocations align each other reducing the intersection angle to zero, the intersection energy decreases.

\section*{Appendix 2: Dislocation structure analysis with / without the cell structures}
\begin{figure}[h!]
  \centering \includegraphics[width=1.0\linewidth]{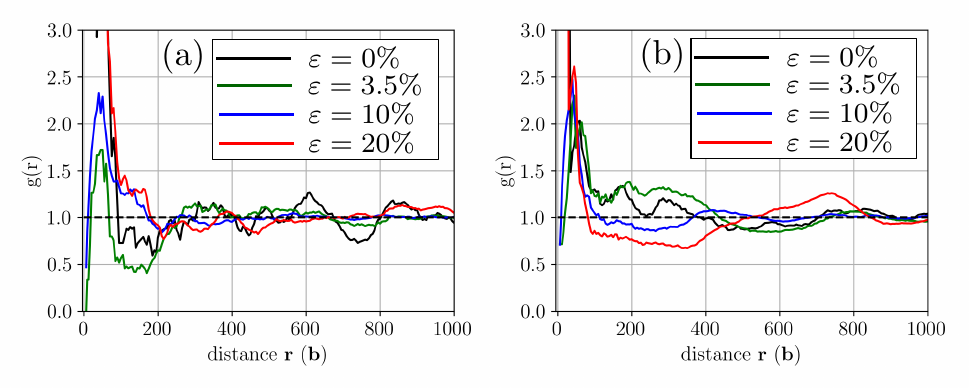}
  \caption{Pair correlation analysis for the four different strain levels $\varepsilon=0\%$, $3.5\%$, $10\%$ and $20\%$ of the (a) no cell and the (b) cell wall systems}
   \label{fig:pair_evol}
\end{figure}
A. Figure~\ref{fig:pair_evol} shows pair correlation analysis of the simulations (a) without the cells and (b) with the cell walls at strains $\varepsilon=0\%$, $3.5\%$, $10\%$ and $20\%$. We observe the simulation without the cells starting with the screw array shows diminishing of correlations as strain increases, while the system with the cell walls starting with the pseudo array exhibits pronounced correlations as strain increases.

\begin{figure}[h!]
  \centering \includegraphics[width=0.9\linewidth]{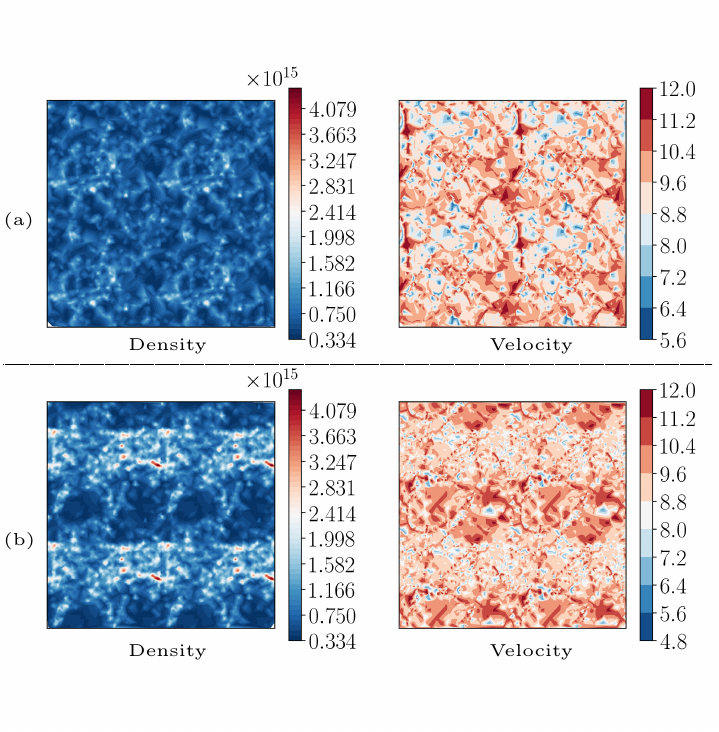}
  \caption{Density and velocity contour maps in the two systems: (a) no cellular structure and (b) cellular structure.}
   \label{fig:dns_contour}
\end{figure}
A. Figure~\ref{fig:dns_contour} shows contours of dislocation densities (left) and velocities (right) (a) without the cells and (b) with the cell walls, respectively, shown in top view. Closely packed dislocations yield low dislocation velocities. With the cells, the dense regions group around the cell walls, while without the cells, such regions are uniformly distributed over the domain.

\begin{figure}[h!]
  \centering \includegraphics[width=0.9\linewidth]{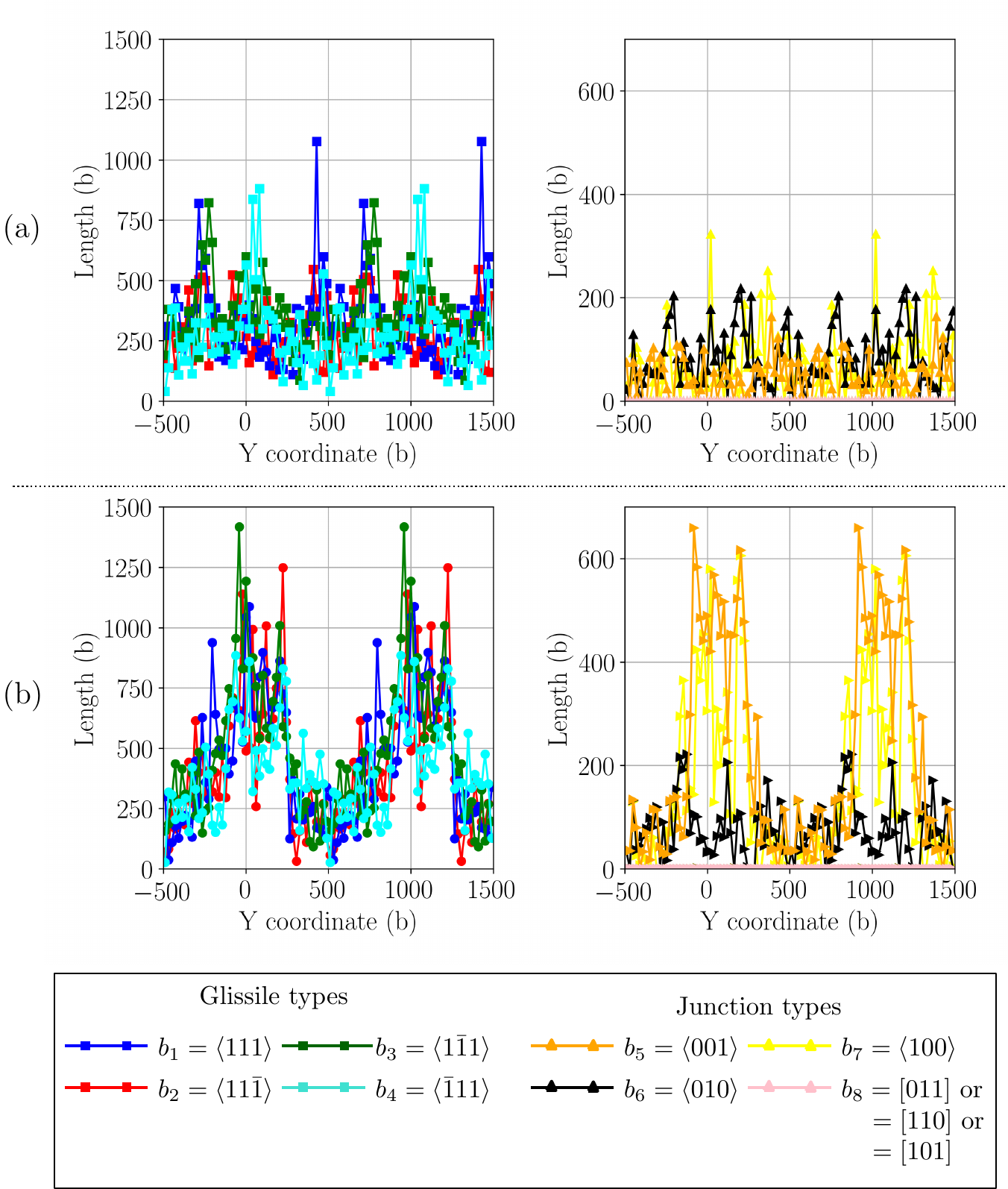}
  \caption{Distributions of dislocations with different types of Burgers vector in the two systems: (a) no cellular structure and (b) cellular structure.}
   \label{fig:btype}
\end{figure}

A. Figure~\ref{fig:btype} shows distributions of dislocation Burgers types and its lengths along the [010] axis (a) without the cells and (b) with the cell walls, respectively. The glissile Burgers types are shown on the left, and junction Burgers types are shown on the right. 
Without the cells, all Burgers types are homogeneously distributed over the domain.
With the cell walls, all the glissile Burgers types are equally distributed and mainly closely packed around the cell walls.
For the junction Burgers types, the two types $b_5$ and $b_7$ are major components of the cell walls.
Those two junction Burgers vectors can be formed between the non-coplanar dislocations as follows:
\begin{eqnarray}
  b_5\langle001\rangle=b_1+b_2 \mbox{ or } b_3+b_4 \\
  b_7\langle100\rangle=b_1+b_4 \mbox{ or } b_2+b_3
\label{eq:jtypes}
\end{eqnarray}
These observations show that the cell walls are composed of coplanar glissile dislocations ($b_1$ and $b_3$ / $b_2$ and $b_4$) hold by junctions created by reactions between non-coplanar glissile dislocations.

\begin{figure}[h!]
  \centering \includegraphics[width=0.7\linewidth]{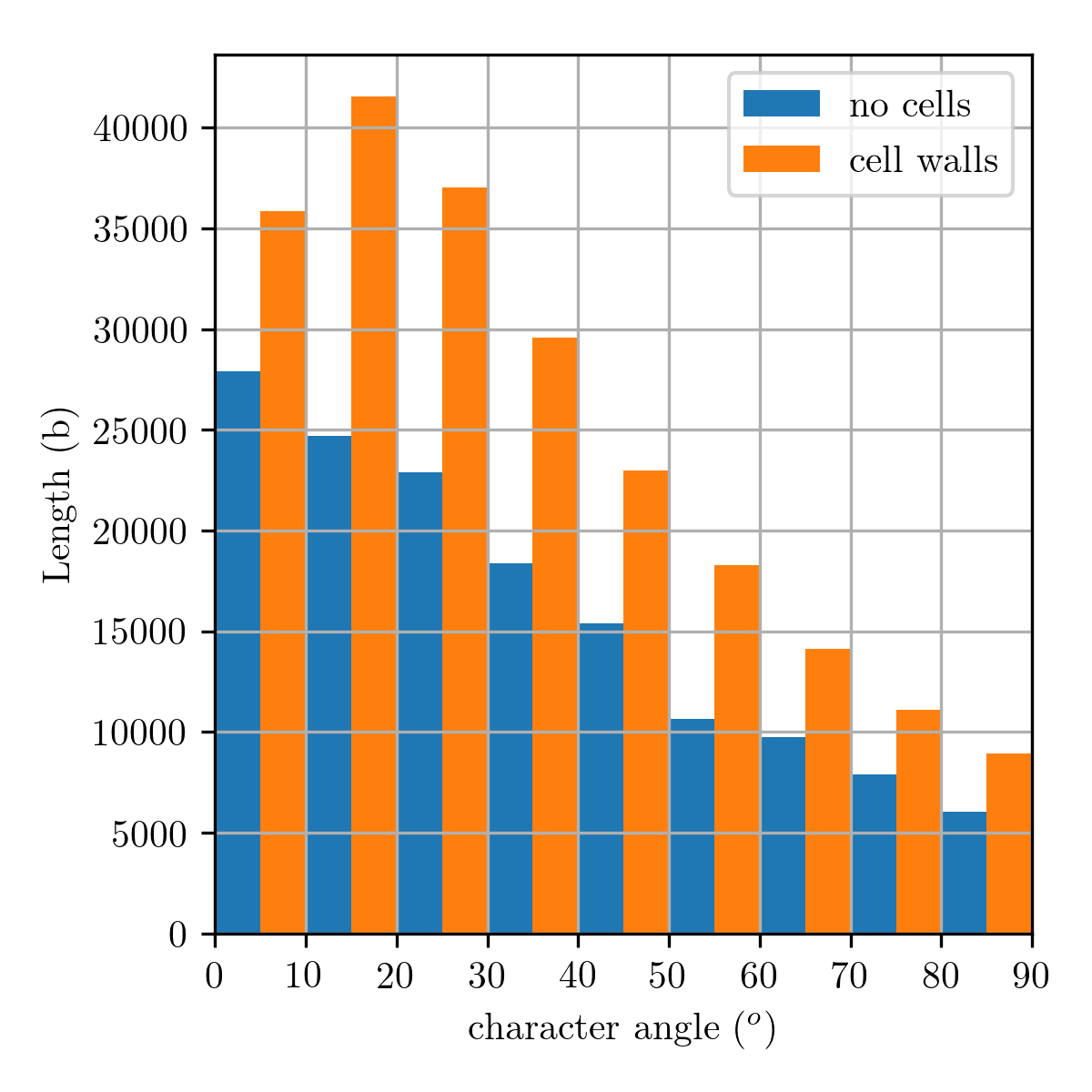}
  \caption{Distributions of character angles of dislocations in the two systems: no cellular structure (no cells) and cellular structure (cell walls).}
   \label{fig:character}
\end{figure}

A. Figure~\ref{fig:character} shows distributions of character angles of dislocations without the cells and with the cell walls shown in blue and orange bars, respectively.
As shown in other DDD simulations of BCC materials~\cite{Arsenlis2003SimulationsOT}, our simulation without the cellular structures, a majority of character angles is screw ($\theta<10^{\circ}$). With the cell walls, the low angle mixed characters ($10^{\circ}\leq\theta<20^{\circ}$) are commonly found.


\end{document}